\documentclass[onecolumn,secnumarabic,amssymb, aps, pra]
{revtex4-1}
\usepackage{graphicx}
\usepackage{epsfig}
\usepackage{xcolor}
\usepackage{inma2ark}
\usepackage{form10ark}
\usepackage{afig1ark}
\usepackage{epic}\usepackage{eepic}
\usepackage{amsmath}

\def\Label#1{\ifbozze\qquad\Red{#1}\fi\Label{#1}}

\def\k(#1){|#1\rangle}
\def\B(#1){{\textbf{#1}}}
\def\C(#1){\mathcal {#1}}
\def\lb{\lambda}
\def\Ld{\Lambda}
\def\M(#1){\mathbb{#1}}
\def\mn{\medskip\noindent}
\def\ov{\overline}
\def\q{\quad}
\def\TT{{\hbox{\tiny T}}}
\def\bm{\begin{bmatrix}} 
\def\em{\end{bmatrix}} 
\def\vq{\;,\quad}
\newcommand{\ba}{\begin{array}}
\newcommand{\ea}{\end{array}}
\newcommand{\beq}{\begin{equation}}
\newcommand{\eeq}{\end{equation}}
\newcommand{\beqa}{\begin{eqnarray}}
\newcommand{\eeqa}{\end{eqnarray}}
\newcommand{\tit}{\textit}

\newtheorem{prop}{Proposition}

\def\BH{{\bf H}}\def\BS{{\bf S}}\def\Bh{{\bf h}}\def\Bs{{\bf s}}
\newif\ifbozze
\bozzefalse
\doppiafalse
\def\qui(#1){\relax}
\def\Label#1{\ifbozze\qquad\Red{ #1}\fi\label{#1}}
 \newcommand*\Red[1]{\textcolor{red}{#1}}
     \newcommand*\Blue[1]{\textcolor{blue}{#1}}
    
\def\nn{\nonumber}    
    
  \newcount\xF     \newcount\yF    
\newcount\xD     \newcount\yD     \newcount\lD   
\def\Z=#1{\ifcase#1 \global\let\harat=\displaystyle        % 0
                \or \global\let\harat=\textstyle          % 1
                \or \global\let\harat=\scriptstyle      % 2
                \or \global\let\harat=\scriptscriptstyle %3
                \fi } 
		 
\def\Fig(#1){\global\advance\nfig by1%
Fig.~\hskip-2pt\the\nfig%
       \ifbozze\margine{{\large\bf F.}\q#1}\fi%
  \begin{figure}[htb]\setlength{\unitlength}{0.500mm}%
     \Label{#1} %\Label{\the\nfig}
      \input{./DD/#1} \end{figure}%
}
\def\Figb(#1){\global\advance\nfig by1%
Fig.~\hskip-2pt\the\nfig%
       \ifbozze\margine{{\large\bf F.}\q#1}\fi%
  \begin{figure}[b]\setlength{\unitlength}{0.500mm}%
     \Label{#1} %\Label{\the\nfig}
      \input{./DD/#1} \end{figure}%
}       
\def\Figt(#1){\global\advance\nfig by1%
Fig.~\hskip-2pt\the\nfig%
       \ifbozze\margine{{\large\bf F.}\q#1}\fi%
  \begin{figure}[t]\setlength{\unitlength}{0.500mm}%
     \Label{#1} %\Label{\the\nfig}
      \input{./DD/#1} \end{figure}%
}
\def\fig(#1){Fig.\hskip1mm\ref{#1}}
   
\begin{document}

\title{From Hamiltonians to complex symplectic transformations}

\author{Gianfranco Cariolaro and Gianfranco Pierobon}%
\email[REVTeX Support: ]{gianfranco.pierobon@unipd.it}
\affiliation{Universit\`a di Padova, Padova, Italy}
\date{\today}%

%\tableofcontents

\begin{abstract}
\ev{Abstract} Gaussian unitaries are specified by a second order polynomial in
the bosonic operators, that is, by a quadratic polynomial and a linear term.
From the Hamiltonian other equivalent representations of the Gaussian unitaries
 are obtained, such as Bogoliubov and real symplectic transformations. 
 The paper develops an alternative representation, called complex
 symplectic transformation, which is more compact and is comprehensive of both
Bogoliubov and real symplectic transformations. Moreover, it  has other advantages.
One of the main results of the theory, not available in the literature, is that the  final displacement \qui(A1) is not simply
given by the linear part of the Hamiltonian, but depends also on the quadratic part. In particular, it is shown that by combining squeezing and rotation,
it is possible to achieve a  final displacement with an arbitrary amount.\\
{\bf}\\
{\bf Symbols and terminology }\\
{\bf}\\
\def\arraystretch{1.1}
\begin{tabular}{ll}
  $:=$               & equal by definition\\
  $I_{\C(H)}$        &identity operator of $ \C(H)$\\
  $I_n$              &identity matrix of size $n$\\
  $A^*$              &adjoint of operator $A$\\
  $\overline x$	     & complex conjugate of the scalar $x$\\	     
  $A^\TT$            &transpose  of matrix $A$\\
  $\overline A$      &conjugate of the matrix $A$\\
 $\C(H)=\C(H)^{\otimes n}$        & bosonic Hilbert space\\
  $q_i$ and $p_i$               & quadrature operators of the $i\,$th mode\\
  $ a=\bm a_1\\\vdots\\ a_n\\\em\q a_\star= \bm a_1^*\\\vdots\\ a_n^*\\\em$
            &vectors of creation and annihilation operators\\[2mm] 
  $ \xi=\bm a\cr a_\star \\\em=
             \bm a_1\\\vdots\\ a_n\\a_1^*\\\vdots\\ a_n^*\\\em$
            &vector of bosonic operators\\[2mm] 
  $\Omega =\bm I_n&0\cr0&-I_n\\\em $& fundamental symplectic matrix  \\
  $H=\frac12 \xi^*\,\BH \, \xi+\xi^*\Bh$
              & quadratic Hamiltonian\\[2mm]
$(\BH\vl\Bh)$ 
              &  matrix representation of $H$\\
$\BH =\bm A&B\cr\overline B&\overline A\\\em
                \q \Bh= \bm h\cr \overline h\\\em$
              &  \\
$(\BS\vl\Bs)$ 
              & matrix representation of complex symplectic transformation\\

$\BS=\bm E&F\cr \overline F&\overline E\\\em \q
                       \Bs= \bm s\cr \overline s\\\em$ 
                   & \\	
$\BS_0$& real symplectic matrix\\ 
$D(\alpha)$, $\;\alpha\in\M(C)^n$ &
      $n$--mode displacement operator  \\
$R(\phi)$,  $\;\phi$ Hermitian matrix &
      $n$--mode rotation operator  \\ 
$Z(z)$, $\;z$ symmetric matrix    &
           $n$--mode squeeze operator\\  
		  \end{tabular}
		  {\bf}\\
		  {\bf}\\
The special symbol $a_\star $ is introduced to denote the
     column vector of the $n$ creation operators. 
        The reason is that, in our conventions,
    $a^*$ is the conjugate transpose  of the column vector  $a$ 
     and therefore it  
  denotes a row vector. 
  The overline  denotes the complex conjugate. Boldface is used only
  for the matrices $\BH$ and $\BS$ and for the column vectors $\Bh$ and $\Bs$. 
		     		     \end{abstract}

\maketitle

\section{Introduction}

In the past three decades the attention to multimode Gaussian states and
transformations has obtained an increasing interest from both the theoretical and
the application point of view. Concepts and protocols, such as entanglement and
teleportation, initially intended only for discrete quantum systems, have been
extended to continuous variable systems, allowing more efficient implementation
and measurements, in particular in the applications of linear optics. As a
consequence, the characterization of Gaussian states and transformations plays a
fundamental role, as witnessed by the large number of review and tutorial papers
devoted to this topic (see f.i.
\cite{Schu86,Brau05,Ferr05,Wang07,Weed12,Oliv12,Ades14}).

Gaussian states are traditionally characterized by waveforms or by Wigner functions, 
expressed as Gaussian functions of the canonical position and momentum coordinates. 
In this approach, Gaussian transformations are defined as unitary transformations that
preserve the Gaussian nature of the states. In a different approach, adopted in 
this note, suitable definitions are given for Gaussian transformations, and 
Gaussian states are defined as obtained by Gaussian transformations applied to the 
the vacuum state (pure Gaussian states) or, more generally, to thermal states 
(mixed Gaussian states). 

We consider an $n$--mode bosonic system characterized by a Hilbert space $\C(H)=
\C(H)_0^n$ with annihilation operators $a_1,\ldots,a_n$ and creation operators 
$a_1^*,\ldots,a_n^*$, satisfying the usual bosonic commutation 
relations $[\,a_i,a_j
\,]=[\,a_i^*,a_j^*\,]=0$ and $[\,a_i,a_j^*\,]=\delta_{ij}$. In this context
the Gaussian unitaries may have several specifications, as shown in \Fig(DD66):

\IT 1) {\it Hamiltonian specification}, given by a second--order polynomial in the bosonic operators; 
	
 \IT 2) {\it Bogoliubov specification}, based on  Bogoliubov transformations;

\IT  3) {\it real symplectic specification} in the phase space,

\IT 4)  {\it complex symplectic specification} in the phase space.

\ET These  specifications are equivalent in representing the whole
class of Gaussian unitaries in the sense that it is possible to obtain
any specification from the others \cite{CarFU}. 
     
 \noindent The Hamiltonian  specification is supported by
a fundamental theorem \cite{MaRh90}, which states that
a unitary operator $ U=e^{-i H}$,
where the Hamiltonian $H$  is a second--order
polynomial
in the bosonic operators $a_i^*$ and $a_i$, is a Gaussian unitary. Then $H$
 can be handled using  a matrix
  representation $(\BH,\Bh)$, having the structure
\beq
\BH =\bm A&B\cr\overline B&\overline A\\\em \vq \Bh=
     \bm h\cr \overline h\\\em \Label{HUH12}
\eeq
where in the $n$--mode $\BH$ is a $2n\times 2n$ complex matrix and $\Bh$ is
a $2n$ complex vector. 
 From  Hamiltonians one can derive  Bogoliubov
transformations, which usually  are formulated in terms of the
 vectors $a\vl a_\star$ containing the bosonic operators $a_i\vl a^*_i$
and have the form
\beq 
a\;\to\; Ea+F a_\star+s
\Label{HUH13}
\eeq
 where $E\vl F$ are $n\times n$ complex matrices 
and $s$ is an $n$ complex vector.  In this paper  we consider a more efficient approach where  the two bosonic vectors $a$ and $a_\star$ are stored
 in a single vector
$\xi$ of size $2n$, thus obtaining the compact  form
\beq 
\xi\;\to\;\; \BS\,\xi+\Bs\vq \q \xi=\bm a\cr a_\star\\\em
\Label{HUH14}
\eeq
where
$(\BS\vl\Bs)$ have the same structure and dimension as $(\BH\vl \Bh)$ in
(\ref{HUH12})
\beq
\BS =\bm E&F\cr\overline F&\overline E\\\em \vq \Bs=
     \bm s\cr \overline s\\\em\;.\Label{HUH16}
\eeq
The  form  (\ref{HUH14}) is developed in \cite{Ades14} and, although
 achieved with a trivial recast of symbols,  has several advantages
  with respect
to  the traditional Bogoliubov form (\ref{HUH13}), namely:
 1) $\BS$ is directly given by an exponential of $\BH$ 
   as $\BS=e^{-i\Omega\BH}$,
  2)  $\BS$ generates  directly the Bogoliubov matrices $E$ and $F$, 
  as indicated in (\ref{HUH16}), and 3) $\BS$
gives directly the traditional real symplectic matrix $\BS_0$
 of the phase space. \qui(A3)In other words, the compact 
 form ({\ref{HUH14}) provides both
the Bogoliubov transformation and the passage to the phase space. 
For these reasons we call $\BS$ {\it complex symplectic matrix}
and the compact form (\ref{HUH14}) {\it complex symplectic transformation}.

In this paper a particular attention is also devoted to the linear
terms $\Bh$ appearing in the Hamiltonian  
 and $\Bs$ in the complex
symplectic  transformation. For convenience  we call
  $\Bs$  {\it displacement amount} or simply {\it displacement}.
In the review papers the linear term is often ignored by saying that it
may be compensated by local operations. In other cases the 
Hamiltonians and the Bogoliubov transformations are presented as alternative 
representations.\qui(A4)  This may inspire the erroneous idea that the displacements 
are exclusively due to the linear part $\Bh$ of the 
Hamiltonian representation: in symbols $\Bs=\Bh$. On the contrary,
in the deduction of the complex symplectic transformation we have discovered that
the displacement $\Bs$ depends not only on the linear term but also,
in a relevant manner, on the quadratic term. 

The paper is organized as follows. In Section II we introduce the matrix representation $(\BH\vl\Bh)$ of a quadratic Hamiltonian and then we prove
the fundamental result on the complex symplectic transformation, giving
the symplectic pair $(\BS,\Bs)$. The proof is based on a sophisticated 
application of the Hadamard Lemma. In particular we will find that
the complex symplectic matrix $\BS$ depends only  on the quadratic part
$\BH$  of the Hamiltonian, while the displacement amount $\Bs$
depends both on $\BH$  and $\Bh$. We also derive from the complex
 symplectic matrix $\BS$ the  standard
real symplectic matrix $\BS_0$.
In Section III   we apply the theory of complex symplectic transformations 
to the fundamental
Gaussian unitaries (displacement,
rotation, and squeezing) with the main target of illustrating the dependence of the displacement
amount $\Bs$ on both $\BH$ and $\Bh$. In  Section IV we will discuss
the possibility of amplification of the displacement amount
by combining  rotation and squeezing. The appendix collects a few
theoretical  topics related to the complex symplectic  transformations.

\section{Generation of complex symplectic transformations}
\bigskip  
\subsection{Specification of a quadratic Hamiltonian}
A  Hamiltonian $H$  given by a second--order polynomial in the bosonic operators
can be written in the form
\beq
H={1\over2}\sum_{r,s=1}^n[\,A_{rs}a_r^*a_s+\overline A_{rs}a_ra_s^*\,]+{1\over2}
	\sum_{r,s=1}^n[\,B_{rs}a_r^*a_s^*+\overline B_{rs}a_ra_s\,]+
	\sum_{r=1}^n(h_ra_r^*+\overline h_ra_r)\;.\Label{Hes}
\eeq
Collecting the coefficients $A_{rs}$, $B_{rs}$ and $h_r$  
in the matrices $A$, $B$, and in  the column vector $h$, the Hamiltonian takes   the compact form  
\beq
H={1\over2}\xi^*\,\BH\xi+\xi^*\,\Bh\Label{Hcomp}
\eeq
where 
\beq
\xi:=
\begin{bmatrix} a\\ a_\star\\
\end{bmatrix}\quad{\rm with}\quad 
a=
\begin{bmatrix}
a_1\\\vdots\\\ a_n\\
\end{bmatrix}\quad,\quad
a_\star=
\begin{bmatrix} 
a_1^*\\\vdots\\a_n^*\\
\end{bmatrix}\Label{Hmat}
\eeq
and
\beq
\BH= 
\begin{bmatrix}
A&B\\\overline B&\overline A\\
\end{bmatrix}\quad,\quad\Bh=
\begin{bmatrix}h\\ \overline h\\
\end{bmatrix}\;.\Label{Hh}
\eeq
$\BH$ is a $2n\times 2n$ Hermitian matrix and $\Bh$ is a $2n$ complex column 
vector. The pair $(\BH,\Bh)$ gives the {\it matrix representation} of $H$. 
The Hermitian nature of the Hamiltonian implies the conditions
\beq
A=A^*\vq B=B^\TT\;, \Label{AB}
\eeq
that is, $A$ must be Hermitian and $B$ must be symmetric. 

For later use it is convenient to decompose the Hamiltonian  (\ref{Hcomp}) 
into the quadratic and the linear parts, namely
\beq
H=H_q+H_\ell\quad{\rm with}\quad  H_q={1\over2}\xi^*\BH\xi\vq
H_\ell=\xi^*\Bh\;.\Label{ABC}
\eeq

\subsection{The fundamental result}

\def\BH{{\bf H}}\def\BS{{\bf S}}\def\Bh{{\bf h}}\def\Bs{{\bf s}}

\medskip\noindent{\sc Theorem 1.} \tit{A Gaussian unitary $U=e^{-iH}$ with 
$H=\frac12\xi^*\BH\,\xi+\Bh^*\,\xi$ applied to the vector  $\xi$ of
 bosonic operators gives the affine transformation
\beq
e^{iH}\xi e^{-iH}=\BS\,\xi+\Bs\vl\Label{A2}
\eeq
where
\beq
\BS=e^{-i\Omega\BH}\vq \Bs=\Psi\Bh
\eeq
with
\beq\nn
\Omega=\bm I_{n}&0\\0&-I_{n}\\\em
\eeq
($I_n$ is the identity matrix of order $n$)
and
\beq
\Psi=\sum_{n=1}^\infty\frac1{n!}(-i\Omega \BH)^{n-1}(-i\Omega )
=(\BS-I_{2n})\BH^{-1}\;.\Label{Psi}
\eeq
}
\medskip

Note that (\ref{Psi}) holds when the matrix $\BH$ is invertible. However, the 
series can be summed in a closed form also when  $\BH$ is singular, as shown in 
Appendix A.
\medskip

The theorem states the passage from the bosonic representation $(\BH,\Bh)$
to the complex symplectic representation $(\BS,\Bs)$, where $\BS$ is a $2n\times 2n$ 
complex symplectic matrix and $\Bs$ is a $2n$ complex vector. For convenience we 
call $\Bs$  ``displacement'' or ``displacement amount''. 

The structure of the matrix $\BH$ in (\ref{Hh}) is invariant with respect to  
matrix addition and multiplication. Since this is also the structure of the matrix 
$-i\Omega$, it follows that the matrices $\BS$ and $\Psi$ have the same 
structure as $\BH$ and $\Bs$ has the same structure as $\Bh$. Namely, 
\beq
 \BS=\bm E&F\\\overline F&\overline E\\\em
\vq\Psi =\bm P&Q\\\ov Q&\ov P\\\em
\vq \Bs=\bm s\\\overline s\em\;.\Label{KB8}
\eeq

In terms of blocks (\ref{A2}) reads
\beq
e^{iH}\bm a\\ a_\star\\\em\,e^{-iH}=
\bm E&F\\\overline F&\overline E\\\em
   \bm a\\a_\star\\\em+\bm s\\\overline s\\\em\Label{Bogm}
\eeq
that is,
\begin{align}\nn
&e^{iH}ae^{-iH} =E\,a+F\,a_\star+s\\
&e^{iH}a_\star e^{-iH} =\overline Ea_\star+\overline F\,a+\overline s\\
\end{align}
with
\beq
s= P\,h+Q\,\ov h\;. \Label{HH4}
\eeq

The complex symplectic matrix $\B(S)$ satisfies the condition
\beq
\B(S)\Omega\B(S)^*=\Omega\;. \Label{BL5}
\eeq
This equation is related to the fact that the Gaussian transformation preserves the
above cited bosonic correlation relations. As a matter of fact, these relations
applied to the vector bosonic operator $\xi$ may be expressed in the compact form
$[\,\xi,\xi^*\,]=\Omega$, from which for the output bosonic operator $\B(S)\xi+
\B(s)$ one gets
\beq\nn 
[\,\B(S)\xi+\B(s),\xi^*\B(S)^*+\B(s)^*\,]=\B(S)[\,\xi,\xi^*\,]\B(S)^*=\B(S)\Omega
\B(S)=\Omega\;,
\eeq
so that the bosonic correlation relations are preserved.

Two particular cases are of interest:

\IT 1) $\Bh=0$: the Hamiltonian reduces to  the quadratic part ($H=H_q$)
and one gets the linear transformation
\beq\nn 
e^{iH_q}\,\xi\,e^{-iH_q}=\BS\;\xi\;.
\eeq

\IT 2) $\BH=0$: the Hamiltonian reduces to  the linear part ($H=H_\ell$) and one 
gets the simple displacement
\beq
   e^{i H_\ell}\,\xi\,e^{-i H_\ell}=\xi-i\,\Omega\,\Bh\;.
  \Label{A2B}
\eeq

\smallskip
\noindent{\bf Remark.} 
In this paper we do not make any reference to Lie groups and 
algebras
\cite{Hall03}, but it would be easy to verify that the matrix $-\Omega\B(H)$,
with $\B(H)$ as in (\ref{Hh}) satisfying conditions (\ref{AB}) form a Lie algebra,
which through $\B(S)=e^{-i\Omega\B(H)}$ generates the complex symplectic Lie group
of the matrices $\B(S)$ satisfying the condition (\ref{BL5}).
 
 \def\lb{\lambda}
\let\T=\TT
\def\mn{\medskip\noindent}

\subsection{Proof of Theorem 1}
 
With the Hamiltonian $H$ decomposed as in (\ref{ABC})
one gets the commutation relations
\beq
[\,iH_\ell,a_k\,]=i[\,a^*h,a_k\,]=-ih_ke_k\quad,\quad
[\,iH_\ell,a^*_k\,]=i[\,a^\T\ov h,a^*_k\,]=ih_k^*e_k\Label{pi9}
\eeq
with $e_k$ a $n$--vector with zero entries except a 1 entry in the $k$--th place.
In a compact form
\beq
[\,iH_\ell,\xi\,]=-i\bm h\\-\ov h\\\em=-i\Omega\Bh\;.\Label{pi10}
\eeq
Moreover one gets \def\nn{\nonumber}
\begin{eqnarray}
&[\,iH_q,a_k]&={i\over2}[\,\xi^*\BH\xi,a_k\,]={i\over2}[\,a^*Aa+a^*Ba_*+
			a^\T\overline Ba+a^\T\overline Aa_*,a_k\,]\nn\\	
		%&={i\over2}\{[\,a^*,a_k\,]Aa+[\, a^*Ba_*,a_k\,]+
		%	a^\T\overline A[\,a_*,a_k\,]\}\cr	
& &		={i\over2}\{[\,a^*,a_k\,]Aa+[\, a^*,a_k\,]Ba_*+a^*B[\,a_*,a_k\,]
			+a^\T\overline A[\,a_*,a_k\,]\}\nn\\
& &	=-{i\over2}\{e_kAa+e_kBa_*+a^*Be_k^\T
			+a^\T\overline Ae_k^\T\}\nn\\
	& &	=-{i\over2}\{e_kAa+e_kBa_*+e_kB^Ta_*
			+e_kA^*a\}=-i(e_kAa+e_kBa_*)\Label{p10}
\end{eqnarray}
and in  compact form 
\beq
[\,iH_q,a\,]=-i(Aa+Ba_*)\;.\Label{pi12}
\eeq
A similar computation gives
\beq
[\,iH_q,a_*\,]=i(\ov Aa_*+\ov Ba)\Label{pi13}
\eeq
and combining (\ref{pi12}) and (\ref{pi13}) yields
\beq
[\,iH_q,\xi\,]=i\bm-A&-B\\\ov B&\ov A\\\em\xi=-i\Omega\BH\xi\;.
\Label{pi14}
\eeq
The above relations are used in the Hadamard identity allowing one to write
\beq
e^{iH}\xi e^{-iH^*}=\sum_{n=0}^\infty{1\over n!}D_n\;,\Label{pi15}
\eeq
where the operator vectors $D_n$ are evaluated by setting $D_0=\xi$ and recursively 
$D_{n+1}=[\,iH,D_n\,]$. Recursion gives
\beq
D_n=(-i\Omega\BH)^n\xi+(-i\Omega\BH)^{n-1}(-i\Omega\Bh)\;.\Label{pi16}
\eeq
Indeed
\beq
D_1=[\,iH,\xi\,]=[\,iH_q,\xi\,]+[\,iH_\ell,\xi\,]=-i\Omega\BH\xi-i\Omega\Bh
\Label{pi17}
\eeq
and (\ref{pi16}) holds true for $n=1$. Moreover, provided that (\ref{pi16}) holds 
true for $n$,
\beq
D_{n+1}=[\,iH,D_n\,]=[\,iH,(-i\Omega\BH)^n\xi\,]=(-i\Omega\BH)^n[\,iH,\xi\,]
=(-i\Omega\BH)^nD_1\;.\Label{pi18}
\eeq
It follows 
\begin{eqnarray}
&e^{iH}\xi e^{-iH^*}
&=\sum_{n=0}^\infty{1\over n!}(-i\Omega\BH)^n\xi
	+\sum_{n=1}^\infty{1\over n!}(-i\Omega\BH)^{n-1}(-i\Omega\Bh)\nn\\
& &=e^{-i\Omega\BH}\xi+\sum_{n=1}^\infty{i^n\over 
n!}(\Omega\BH)^{n-1}(-i\Omega\Bh)=e^{-i\Omega\BH}\xi+\Bs\;.
\end{eqnarray}

\subsection{The fundamental Gaussian unitaries}	
		   
 To proceed it is convenient to introduce the 
 fundamental Gaussian unitaries (FUs). These unitaries were formulated for multimode systems 
 by Ma and Rhode  \cite{MaRh90}, through the following definitions:
 
\IT 1) {\it Displacement operator}
\beq
D(\alpha):=\E^{ \alpha^\TT a_\star\,\,-\alpha^*\, a} 
\vq \qq\alpha=[\alpha_1,\ldots,a_n]^\TT \in \M(C)^n
\Label{U22}
\eeq
which is the same as the Weyl operator.
\IT 2) {\it  Rotation operator}
\beq
R(\phi):=\E^{\I\, a^*\phi \, a}\vq \q\q\phi
 \q\hb{$n\times n$ Hermitian matrix}\;.
\Label{U24}
\eeq
\IT 3) {\it  Squeeze operator} \cite{squeeze}
\beq
 Z(z):=\E^{\met\left[\,( a^* \,z\,{ a_\star}-a^\TT\,z^*\,a)\right]} 
\vq \!\!z\q \hb{$n\times n$  symmetric  matrix}\;.
\Label{U26}
\eeq
\ET

 The importance of these operators, illustrated in  \Figb(DD5), is established by the following:
 
\medskip\noindent{\sc Theorem 2} \cite{MaRh90}. \tit{
The most general Gaussian unitary is given by the combination of the
 three fundamental Gaussian unitaries $D(\alpha)$, $Z(z)$,
  and $R(\phi)$, cascaded in any arbitrary order, that is, 
\beq
Z(z)\,D(\alpha)\,R(\phi)\vq
 R(\phi)\,D(\alpha)\,Z(z), \q\hb{etc.}\Label{KK}
 \eeq
}

It is interesting to remark that the above definitions can be obtained
from  the matrix representation of the
Hamiltonian
\beq\nn 
\BH =\bm A&B\cr\overline B&\overline A\\\em\vq \Bh=
     \bm h\cr \overline h\\\em\q\to\q H={1\over2}\xi^*\,\BH\xi+\xi^*\,\Bh\;.
\eeq
by setting
 to zero two of the submatrices $A\vl B\vl h$. Specifically
 (see Appendix B):
\IT 1) {\it Displacement} with $A=B=0$ and $h\in\M(C)^n$
\beq
H=\xi\;\Bh=h^\TT\;a_\star+h^*\;a\q\to\q D(\alpha)=e^{-i H}=\E^{ \alpha^\TT a_\star\,\,-\alpha^*\, a} 
\vq \qq\FB{\alpha=-i h }
\Label{U22A}
\eeq
\IT 2) {\it Rotation} with $B=0\vl h=0$ and  $A$ $n\times n$ Hermitian
\beq
H=\met(a^*\,A\,a+a^\TT\,\ov A\,a_\star)
         \q\to\q R(\phi)=e^{-i H}=e^{i\, a^*\phi \, a}\vq \q\q\FB{\phi=-A}
\Label{U24A}
\eeq
\IT 3) {\it Squeezing} with $A=0\vl h=0$ and $B$ $n\times n$ symmetric
\beq
H=\met(a^*\,B\,a+a^\TT\,\ov B\,a_\star)  \q\to\q  Z(z):=\E^{\met\left[\,( a^* \,z\,{ a_\star}-a^\TT\,z^*\,a)\right]} \vq\q\q \FB{z=-iB}\;.
\Label{U26A}
\eeq

We consider in particular the displacement. In the above definition
  $D(\alpha)$ is formulated in  terms
of the bosonic vectors $a,a_\star$. But for the interpretation of
Theorem 1 is is convenient to give a formulation in terms of the
single vector $\xi$, as
\beq
 \C(D)_Y(\xi):=e^{\xi^*\,\Omega\,Y}\ {\rm with}\ Y=\bm y\\ \ov y\\\em
 \ ,\ y\in\M(C)^n\;.
  \Label{EE}
\eeq
This definition ensures that $\C(D)_Y(\xi)$ provides the shift of $Y$
 indicated in the symbol
\beq
 \C(D)^*_Y(\xi)\;\xi\;\C(D)_Y(\xi)=\xi+Y\;.
 \Label{A2D}
\eeq
 On the other hand, the standard form  provides the transformations
\beq\nn 
D^*(\alpha) \;a\;D(\alpha) \;=\;a+\alpha\vq
    D^*(\alpha) \,a_\star\,D(\alpha) =
       a_\star+\ov\alpha
\eeq
which are equivalent to
\beq  
D^*(\alpha) \;\xi\;D(\alpha) =\xi+\bm\alpha\\\ov\alpha\\\em\;. \Label{A2F}
\eeq
Comparison of (\ref{A2D}) and  (\ref{A2F}) gives
\beq
	 Y=\bm y\\ \ov y\\\em=\bm \alpha\\\ov\alpha\\\em\;.\Label{LL8}
\eeq
 
\subsection{Interpretation  of the fundamental result}

The complex symplectic transformation  $\BS\,\xi+\Bs$ can be expressed
in two different ways in terms of the two transformations:

\bu the linear transformation $\C(S)_{\BH}(\xi)=e^{-i\Omega \BH}\,\xi$,
with input $\xi$ and output $\BS\,\xi$,

\bu a displacement $\C(D)_Y(\xi)$ with input $\xi$ and output $\xi+Y$,
where $Y$ is a $2n$ complex vector as in (\ref{LL8}).

Then the transformation  $\BS\,\xi+\Bs$ can be expressed by

\IT 1) the  linear transformation  $\C(S)_{\BH}(\cdot)$ followed by the 
   displacement $\C(D)_s(\cdot)$, giving
\beq\nn 
\C(D)_s(\C(S)_{\BH}(\xi))=\C(D)_y(\BS\,\xi)=\BS\,\xi+\Bs
\eeq
or
\IT 2) the displacement $\C(D)_Y(\cdot)$  with  $Y=\BS^{-1}\Bs$ followed
by the  linear transformation  $\C(S)_{\BH}(\cdot)$, giving
\beq\nn 
\C(S)_{\BH}(D_Y(\xi))=\C(S)_{\BH}(\xi+Y)=
e^{-i\Omega \BH}\,\xi+e^{-i \Omega \BH}\,Y=
   e^{-i \Omega \BH}\,\xi+\Bs\;.
\eeq

The  two equivalent interpretations are illustrated in \Fig(CS176). Note that the linear
transformation is common to both cascades, while the displacement is different.

 \subsection{The complex symplectic matrix of a general Gaussian unitary}
 
For a general Gaussian unitary (see Theorem 2) it is possible to
calculate the complex symplectic matrix $\BS$ in a closed form.

\medskip\noindent{\sc Theorem 3.}\tit{ For the general Gaussian unitary
$R(\phi)Z(z)D(\alpha)$ the complex symplectic matrix  is given by
\beq
\BS=\bm\cosh (r)\;e^{i\phi}&\sinh (r)\; e^{i\theta}e^{-i\phi^\TT}\\ 
\sinh (r^\TT)\;e^{-i\theta^\TT} e^{i\phi}& \cosh (r^\TT)\;e^{-i\phi^\TT}\\\em
\Label{Ss}
     \eeq
     where the squeeze matrix is decomposed in the polar form
      $z=r\;e^{i\theta}$.
     }
     
\ET{\it Proof}\q
Consider two $n$--mode Hamiltonians $H_{sq}$  and $H_{rot}$ given by
the matrix representations
\beq\nn  
\BH_{sq}=\bm0&i z\\ -i \ov z&0\\\em\quad,\quad
{\bf H}_{rot}=-\bm\phi&0\cr0&\phi^\TT\em 
\eeq
where $z=r\,e^{i\theta}$ is the squeeze matrix and $\phi$ the rotation matrix.
Considering that $\BH_{disp}=0$,
the global symplectic matrix is obtained as
\beq
\BS=\BS_{sq}\BS_{rot}=e^{-i\;\Omega\,\BH_{sq}}e^{-i\;\Omega\,\BH_{rot}}\;.
\Label{Z14A}
\eeq
Now the evaluation of  $\BS_{rot}$ is immediate
\beq
\BS=e^{-i\Omega\BH_{rot}}
=\bm e^{i\phi}&0\\0&e^{-i\phi^\TT}\em\;.
\Label{DD4}
\eeq 
For the evaluation of $\BS_{sq}$ we use the general formula
\beq
\BS_{sq}=\bm E_{sq}& F_{sq}\\
            \ov F_{sq}& \ov E_{sq} \\\em
	    \eeq
where $E_{sq}$ and $ F_{sq}$ are evaluated in \cite{MaRh90} and read	 
\beq 
     E_{sq}=\cosh(r)\vq F=\sinh(r)\,e^{i\theta}\;.
     \eeq
Then combination of the above results gives (\ref{Ss}).

 \subsection{Relation with  real symplectic transformations}

Real symplectic  transformations refer to the quadrature operators
 arranged in the form $\xi_0=[q,p]^\TT$ and has the same affine structure
 seen for  complex symplectic transformation  (\ref{HUH14}), namely
 \beq
 \xi_0\q\to\q \B(S)_0 \,\xi_0+\Bs_0\;.
\Label{HH30}
\eeq
The relation between the two affine transformations is easily obtained
considering that the quadrature operators are related to the bosonic operator as
$q_i=(a_i+a_i^*)/\sqrt2$, $ p_i=- i\,(a_i-a_i^*)/\sqrt2$,
and in compact form
 \beq
       \xi_0 =L\,\xi \q\hbox{\rm with}\q
   L=\frac1{\sqrt2} \bm I_n &I_n\cr -i \,I_n&i \,I_n \\\em\;.
     \Label{C8}
\eeq
Then, relation  $ \xi\;\to\; \BS \,\xi +\Bs$ gives (\ref{HH30})
with
\beq
 \BS_0=L\;\BS\;L^*  \vq \Bs_0=L\,\Bs\;. \Label{C9}
 \eeq

Now it easy to see that the matrix $S_0$ and the vector $s_0$ are real.	
In fact, (\ref{C9}) gives explicitly 
\beq
\BS_0=\bm \Re( E+ F)&-\Im(E-F)\cr
		           \Im(E+F)&\Re( E- F)\\\em\vq
			    \Bs_0=\sqrt2\bm \Re\, s\cr \Im\, s \\\em
			   \Label{C9B}
 \eeq
From (\ref{C9}) we find the symplectic condition for the matrix $\BS_0$
\beq
 \BS_0\,\Omega_0\,\BS_0^\TT= \Omega_0 \q\hbox{\rm with}\q
\Omega_0= \bm 0&I_n\cr-I_n&0\\\em\;.
   \Label{H5}
\eeq

\smallskip
\noindent{\bf Remark.} In this paper, in order to avoid a proliferation of notations, we adopt the symbols
$\B(S)$ and $\Omega$ which in the literature are usually reserved to the real
symplectic matrices appearing in the Gaussian transformations of canonical
operators, strictly related to the complex symplectic transformations
\cite{Arvi95}.
As a consequence of the unitary similarity (\ref{C9}) the sets of the real
 and complex symplectic groups are isomorphic.

\section{Applications to fundamental Gaussian unitaries}

We apply the previous theory on complex symplectic transformations
 to combinations of  fundamental Gaussian unitaries. 
 More specifically, we develop
the cases:

\IT 1) rotation+displacement,

\IT 2) squeezing+displacement,

\IT 3) squeezing+rotation+displacement (which represents the most general
Gaussian unitaries \cite{MaRh90}\cite{CarQC}).

\ET The results will be expressed  in the general $n$ mode and illustrated
in the single and in the two mode.
We assume that the matrix $\BH$ is not 
singular, so that we can apply the closed--form formula (\ref{Psi}). The detail of the deduction is
given in Appendix C.
\smallskip 

\iffalse
Preliminary we list the $\BH,\Bh)$ and $(\BS,\Bs)$ representations
for each FU.
\qui(B1)
\begin{tabular}{ll ll}
\multicolumn{4}{l}{ 1) Displacement}\\ 
$\BH=0$& $\Bh=\bm h\cr \ov h\\\em$ &$\BS=0$ & $ \Bs=\Bh=\bm h\cr\ov h\\\em=\bm i\alpha\cr-i\ov\alpha\\\em$ \\
%
\multicolumn{4}{l}{ 2) Rotation}\\
$\BH=$& $\Bh=\bm h\cr \ov h\\\em$ &$\BS=0$ & $ \Bs=\Bh=\bm h\cr\ov h\\\em=\bm i\alpha\cr-i\ov\alpha\\\em$ \\
\end{tabular}

   \beq\nn 
   \BS=0\vq  \Bs=\bm h\cr\ov h\\\em=\bm i\alpha\cr-i\ov\alpha\\\em
   \eeq

\IT 2) rotation
   \beq\nn 
   \BS=\bm A&0\cr0&\ov A\\\em =\bm e^{i\phi}&0\cr0&e^{-i\phi}\\\em \vq  \Bs=0
   \eeq
 \IT 3) rotation
   \beq\nn 
   \BS=\bm 0&B\cr \ov B&0\cr\\\em =\bm z&0\cr0&e^{-i\phi}\\\em \vq  \Bs=0
   \eeq
   \fi

\subsection{Rotation+displacement}

The Hamiltonian $H$ is given by the matrix representation 
\beq
{\bf H}=-\bm\phi&0\\0&\phi^\TT\\\em\quad,\quad\Bh=
\bm h\\\ov h\\\em\Label{DD2}
\eeq
where $\phi$ is an $n\times n$ Hermitian matrix and $h$ is an arbitrary complex 
vector of size $n$.

\begin{prop} \Label{P2}
The complex symplectic matrix is given by
\beq
\BS=e^{-i\Omega\BH}
=\bm e^{i\phi}&0\\0&e^{-i\phi^\TT}\em\;.
\Label{DD4}
\eeq
The matrix giving the displacement results in
\beq
\Psi=
     -\bm(e^{i\phi}-I_n)\phi^{-1}&0\\
     0&(e^{-i\phi^\TT}-I_n)\phi^{-\TT}\\\em 
     \qquad \phi\in[-\pi,\pi)^n\Label{DD4A}
\eeq
where $\phi^{-\TT}:=(\phi^{-1})^\TT$.

\end{prop}

\medskip
Expression (\ref{DD4A}) is a special case of the second  of (\ref{KB8}) with $P=-(e^{i\phi}-I_n)\phi^{-1}$ and 
$Q=0$, so that
\beq
s=-(e^{i\phi}-I_n)\phi^{-1}\,h\qquad \phi\in[-\pi,\pi)^n \Label{DD7}
\eeq
which depends on the companion Gaussian unitary (in this case the rotation).

\smallskip
\noindent\tit{Remark: periodicity of $\BS$ and $\Psi$}

\smallskip
\noindent
The matrix $\BS$ is periodic with respect to the phase matrix $\phi$. 
In fact, $e^{i(\phi+2k\pi I_n)}=e^{i\phi}\,e^{i 2\pi k\,I_n}=e^{i\phi}$, 
$k\in \M(Z)$. Then, 
the specification of $\Psi$ can be confined in the
$n$--dimensional period $\C(P)=[-\pi,\pi)^n$.  Also the matrix $\Psi$
should be periodic with the same periodicity, $\Psi(\phi+k\,I_n)=\Psi(\phi)$.
But the expression given by (\ref{DD4A}) is aperiodic and is correct only 
if $\phi$ is confined in the period $\phi\in[-\pi,\pi)^n$. On the other hand,
it is possible to get an unconstrained expression using the identity
\beq
\BH=-\log\left[\,\exp(i\BH\right)] \Label{Dm5}
\eeq
where with $\log$ the principal value of the logarithm should be intended
\cite{Higham}.
As we will see in the single mode, the use of identity (\ref{Dm5}) leads to
cumbersome expressions, so that we prefer the aperiodic
 expressions
like (\ref{DD4A}) with the indication of the validity in a period.
 
\smallskip
\noindent\tit{Example 1.} We discuss (\ref{DD7}) in the single mode, where we get
\beq\nn 
\BH=\bm-\phi&0\\
	0 & -\phi  \\\em
\quad,\quad
-i\Omega\BH=\bm
 i \phi  & 0 \\
 0 & -i \phi  \\\em
\eeq
\beq\nn 
\BS=\bm
 e^{i \phi } & 0 \\
 0 & e^{-i \phi } \\\em\quad,\quad
\BH^{-1}=\bm
 -\frac{1}{\phi } & 0 \\
 0 & -\frac{1}{\phi } \\\em
\eeq
\beq\nn 
\Psi=\bm-\frac{-1+e^{i \phi }}{\phi } & 0 \\
 0 & -\frac{-1+e^{-i \phi }}{\phi } \\\em
\eeq
giving
\beq
P(\phi)= -\frac{-1+e^{i \phi }}{\phi }=- ie^{i\phi/2}
 \frac{\sin(\phi/2)}{\phi/2}\quad,\quad\phi\in[-\pi,\pi)\;. \Label{DD9}
\eeq
This is the ``aperiodic'' solution which  is correct only in the interval
 $\phi\in[-\pi,\pi)$.
To get an unconstrained periodic expression we can use the identity
(\ref{Dm5}), which gives
\beq\nn 
\BH=\bm
 -i\log(e^{-i\phi }) & 0 \\
 0 & i\log(e^{i \phi }) \\\em
 \quad,\quad
\BH^{-1}=\bm
 \frac{i}{\log(e^{-i\phi})} & 0 \\
 0 & \frac{-i}{\log(e^{i\phi })} \\\em
\eeq
\beq\nn 
\Psi_0=\bm
 \frac{i(-1+e^{i \phi})}{\log(e^{-i\phi})} & 0 \\
 0 & \frac{-i(-1+e^{i\phi})}{\log(e^{i\phi})}\\\em
\quad,\quad P_0(\phi)=\frac{-i(-1+e^{i \phi })}{\log(e^{i\phi })}
\eeq

Note that $P_0(\phi)$ is periodic as shown in \Fig(DD189).

\medskip\noindent\tit{Example 2 (Beam splitter).} 
A beam splitter is modeled as a 
two--mode rotation operator with rotation matrix
\beq
\phi=\bm0 & -i \beta\\
 i \beta & 0 \\\em\quad,\quad\beta\in[-\pi,\pi) \Label{L4A}
\eeq
The corresponding matrices $\BH$  and $\BS$ result in
\beq\nn 
\BH=\bm
 0 & i \beta  & 0 & 0 \\
 -i \beta  & 0 & 0 & 0 \\
 0 & 0 & 0 & -i\beta  \\
 0 & 0 & i\beta  & 0 \\\em
\quad,\quad
\BS=\bm\cos\beta & \sin\beta & 0 & 0 \\
 -\sin \beta  & \cos \beta  & 0 & 0 \\
 0 & 0 & \cos\beta  & \sin \beta  \\
 0 & 0 & -\sin \beta  & \cos \beta  \\\em\;.
\eeq
The inverse of $\BH$ is
\beq\nn 
\BH^{-1}=\bm
 0 & \frac{i}{\beta } & 0 & 0 \\
 -\frac{i}{\beta } & 0 & 0 & 0 \\
 0 & 0 & 0 & -\frac{i}{\beta } \\
 0 & 0 & \frac{i}{\beta } & 0 \\\em
\eeq
Hence
\beq\nn 
\Psi=\bm
 -\frac{\sin\beta}{\beta } & \frac{i (\cos \beta -1)}{\beta } & 0 & 0 \\
 -\frac{i (\cos \beta -1)}{\beta } & -\frac{i \sin\beta }{\beta } & 0 & 0 \\
 0 & 0 & \frac{i \sin \beta }{\beta } & -\frac{i(\cos\beta -1)}{\beta } \\
 0 & 0 & \frac{i (\cos\beta-1)}{\beta } & \frac{i\sin \beta }{\beta } \\
\em \vq \beta\in[0,2\pi)
\eeq
\beq\nn 
P=\bm
 -\frac{i\sin\beta}{\beta } & \frac{i (\cos\beta-1)}{\beta } \\
 -\frac{i (\cos\beta-1)}{\beta } & -\frac{i\sin\beta}{\beta } \\
\em\vq \beta\in[0,2\pi)\quad,\quad Q=0\;.
\eeq
Again, $\Psi$ and $P$ are aperiodic, while they should be periodic in $\beta$.
The remedy is the limitation of $\beta$ as in (\ref{L4A}) or the 
application of identity (\ref{Dm5}). But the latter solution gives a cumbersome
expression.

\subsection{Squeezing+displacement}

The Hamiltonian $H$ is given by the matrix representation (see (\ref{U26}))
\beq
\BH=-\bm0&iz\\ -i\ov z&0\\\em\quad,\quad\Bh=
\bm h\\\ov h\\\em
\Label{DD2A}
\eeq
where $z$ is an $n\times n$ symmetric matrix  and $h$ is an arbitrary
complex vector of size $n$. The squeeze matrix must be decomposed
 in polar form as
$z=r\,e^{i\theta}$, where $r$ is positive semidefinite and $\theta$ is Hermitian.

\begin{prop} \Label{P4}
The complex symplectic matrix is given by (see (\ref{Ss}))
\beq
\BS=e^{-i\Omega\BH}=
\bm\cosh (r) & \sinh (r)  e^{i \theta }\\
 \sinh (r^\TT)e^{-i\theta^\TT } & \cosh (r^\TT) \\\em  \Label{HH2} 
\eeq
The matrix giving the displacement results in
\beq
\Psi=\bm  \sinh (r)e^{i\theta }\left(-iz^{-1}\right)
           &(\cosh (r) -I_n)i{\ov z}^{-1}\\
	    (\cosh (r^\TT)-I_n )(-iz^{-1})
	    &\sinh (r^\TT)e^{-i\theta^\TT }i{\ov z}^{-1}\\\em
	    \Label{HH4}
\eeq
\end{prop}

\medskip
Expression (\ref{HH4}) is a special case of the second  of (\ref{KB8}) 
with 
\beq
P= \sinh (r)e^{i\theta }\left(-iz^{-1}\right)\quad,\quad
Q=(\cosh (r) -I_n)i\,{\ov z}^{-1}\;.\Label{DD2C}
\eeq
Hence
\beq
s= \sinh (r)e^{i\theta }\left(-i\,z^{-1}\right)\;h
           +(\cosh (r) -I_n)i\,{\ov z}^{-1}\;\ov h\;.
	   \Label{DD2F}
\eeq
\qui(B1)%fffwhich depend on the companion Gaussian unitary (in this case the squeezing).

\medskip

\noindent\tit{Example 3.}   In the single mode, (\ref{DD2F}) gives
\begin{eqnarray}
&s&=e^{i \theta }\sinh r\left(-i\,z^{-1}\right)\,h
           +(\cosh r -1)i\,{\ov z}^{-1}\;\ov h\nn\\
&  &=-i\frac{\sinh(r)}{r}\;h + 
	    i\frac{\cosh(r)-1}{re^{-i\theta}}\; \ov h\;.\Label{DD2FB}
\end{eqnarray}
where $h=i\alpha$ (see (\ref{U22A})).

The plot of $|s|$ as a function of $r$ and as a function of $\theta$ 
and $h=1$ is shown in \Fig(DD190).

\subsection{Squeezing+rotation +displacement}

Consider two $n$--mode Hamiltonians $H_{sq}$  and $H_{rot}$ given by
the matrix representations
\beq\nn  
\BH_{sq}=\bm0&i\, r\;e^{i\theta}\\ -i\, 
                 e^{-i\theta^\TT}\,r^\TT &0\\\em\quad,\quad
{\bf H}_{rot}=-\bm\phi&0\cr0&\phi^\TT\em 
\eeq
and a general linear term
\beq\nn 
\Bh=\bm h\\\ov h\\\em\;.
\eeq
This specifies the most general Gaussian unitary \cite{MaRh90}.
The global symplectic matrix is obtained as
\beq
\BS=\BS_{sq}\BS_{rot}=e^{-i\;\Omega\,\BH_{sq}}e^{-i\;\Omega\,\BH_{rot}}
\Label{Z14}
\eeq
and reads on (see ({\ref{Ss}))
\beq
\BS=	
\bm\cosh (r)\;e^{i\phi}&\sinh (r)\; e^{i\theta}e^{-i\phi^\TT}\\ \sinh (r^\TT)\;
e^{-i\theta^\TT} e^{i\phi}& \cosh (r^\TT)\;e^{-i\phi^\TT}\\\em\;.
\Label{HT4}
\eeq
But we want to obtain this expression starting from a single Hamiltonian.
Note that in general
\beq
e^{-i\;\Omega\,\BH_{sq}}e^{-i\;\Omega\,\BH_{rot}}
\neq e^{-i\;\Omega\,\BH_{sq}-i\;\Omega\,\BH_{rot}}\;.\nn
\eeq
The single Hamiltonian is given by (see \cite{MaRh90})
\beq
\BH=i\Omega \log(\BS)\;. \Label{PQ12}
\eeq

For the evaluation of  $\log(\BS)$ we can use the standard methods of
calculation of a function of a matrix \cite{Higham}. Then we apply (\ref{Psi}) 
to evaluate the matrix $\Psi$ and (\ref{HH4}) to evaluate 
 the displacement $\Bs=\Psi\Bh$.
	   
\medskip	   

\noindent\tit{Example 4.} We develop the above procedure in the single mode.
For the evaluation of the $2\times2$ matrix $\BH$,  according to (\ref{PQ12}),
we can use the Sylvester interpolation method \cite{CarFU}, which gives
\beq
  \BH=d_0\,I_2+ d_1\,i\,\Omega\,\BS:=\bm A&B\\\ov B&\ov A\\\em
    \Label{M39}
\eeq
The coefficient  $d_0$ and $d_1$ are given by
\beq
d_{0}=\frac{\lb_+^2+1}{\lb_+^2-1}\log(\lb_+)\quad,\quad  
d_{1}=\frac{2\lb_+}{\lb_+^2-1}\,\log(\lb_+)\quad\quad|\Re(E)|\neq1
   \Label{N13}
\eeq
where $\lb_{\mp}$ are the eigenvalues of $\BS$, which result in
\beq\nn 
\lb_-= 
    \Re(E)- \sqrt{\Re(E)^2-1}\quad,\quad \lb_+= 
    \Re(E)+\sqrt{\Re(E)^2-1}
\eeq
and
\beq\nn 
E=\cosh(r) e^{i\phi}\quad,\quad F=\sinh(r)e^{i(\theta-\phi)}\quad,\quad
L=\sqrt{\Re(E)^2-1}\;.
\eeq
We find
\beq\nn 
A=\frac{\Im(E)}{2L}\,\log\frac{(\Re(E)-L}{\Re(E)+L}\quad,\quad
B=-i\, \frac{F}{2 L}\,\log\frac{\Re(E)-L}{\Re(E)+L}
\Label{M40}
\eeq
Hence
\beq\nn 
\Psi=(\BS-I_2)\BH^{-1}=
\bm\frac{2 i (E+1)  |F|^2 L   (\Re(E)-1)}{(X-Y) \left[\left[\left| E\right| ^2-1\right]^2- |F|^2 \Im(E)^2\right]} & \ \frac{2 i  F L^* \left[-(E-1) \left| E\right| ^2+E+ |F|^2 (E-\Re(E))-1\right]}{\left[X^*-Y^*\right] \left[\left[\left| E\right| ^2-1\right]^2- |F|^2 \Im(E)^2\right]} \\
 \ \frac{2 i L  F^* \left[E \left[E^*\right]^2-(E+1) E^*+ |F|^2 (E-\Re(E))+1\right]}{(X-Y) \left[\left[\left| E\right| ^2-1\right]^2- |F|^2 \Im(E)^2\right]} & -\ \frac{i  F \left[\left| E\right| ^2+E+\left[E^*-1\right] E^*-2\right]  F^* L^*}{\left[X^*-Y^*\right] \left[\left[\left| E\right| ^2-1\right]^2- |F|^2 \Im(E)^2\right]} \\\em
\eeq
 with
\beq\nn 
 X=\log(\Re(E)-L)\vq Y=\log(\Re(E)+L)\;.
\eeq
 In particular
\begin{eqnarray}
&P&=\frac{2 i (E+1)  |F|^2 L   (\Re(E)-1)}{(X-Y) \left[\left[\left| E\right| ^2-1\right]^2- |F|^2 \Im(E)^2\right]}\\
&Q&=\ \frac{2 i  F L^* \left[-(E-1) \left| E\right| ^2+E+ |F|^2 (E-\Re(E))-1\right]}{\left[X^*-Y^*\right] \left[\left[\left| E\right| ^2-1\right]^2- |F|^2 \Im(E)^2\right]}
\end{eqnarray}
The displacement amount $s$ is related to the $h$ amount as in (\ref{HH4}),
that is,
$
s=P\,h+Q\ov h$, where
the coefficients $P$ and $Q$ can be expressed in terms of rotation and
 squeezing parameters $\phi$, $r$, and $\theta$ as 
\beq\nn 
P=-\frac{2 i \Delta \left[1+e^{i \phi } \cosh (r)\right]}{A\,(\cos (\phi ) \cosh (r)+1)}\vq
Q=\frac{2 i \Delta^* e^{i \theta -i \phi } \sinh (r)}{A^*\,(\cos (\phi ) \cosh (r)+1)}
\eeq
with
\beq
\Delta=\sqrt{\cos ^2(\phi ) \cosh ^2(r)-1}\quad,\quad
A= \log \left[\frac{-\Delta+\cos (\phi ) 
\cosh (r)}{\Delta+\cos (\phi ) 
\cosh (r)}\right]\;.
\Label{TT3}
\eeq

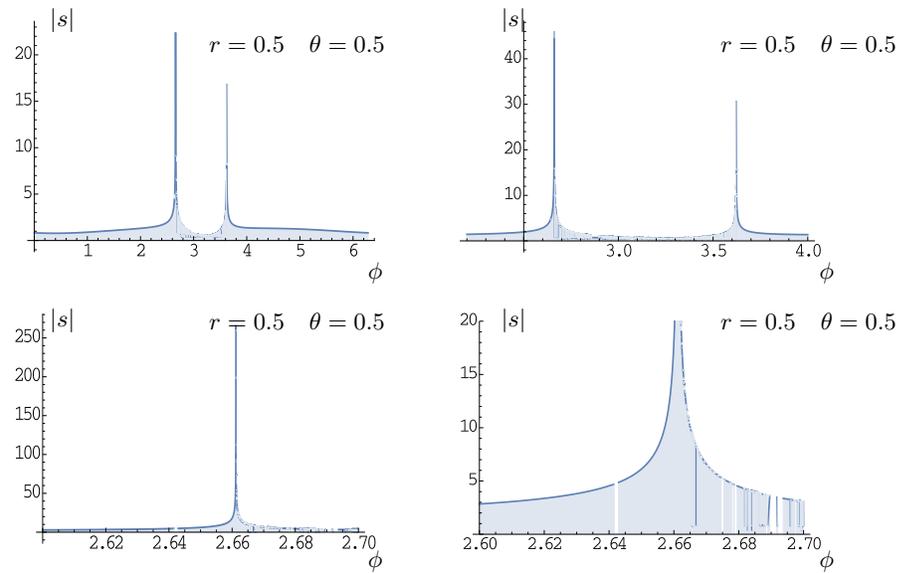
\Figb(DE4010) shows the displacement amount $|s|$ as a function of $\theta$
for four values of $r$ for $\phi=0.3$ . Note 
that the curves increase with $r$ and, for a given $r$, have a maximum for 
$\theta=\pi$.
\def\caplabel{}
 \Fig(DE4014)
 shows the behavior of the displacement amount as a function
of the phase $\phi$. This behavior  exhibits divergences at two points of 
the period $[0,2\pi)$, as shown in Fig. 6. In  fact
  the denominators  of $P$ and $Q$ vanish at such points. 
\newpage
  
  \Fig(DD133) and \Fig(DD192) show other representations of the displacement 
  amount.
 \clearpage 
  \bigskip
  
\ET{\it Alternative evaluation for the single mode} 
\ET In Appendix D we consider a specific evaluation
for the single mode leading to simpler results. We find
$$
\eqalign{
&P=-\frac{i \sinh (T) \left(-e^{i \phi } \cosh (r) (\cosh (T)+1)+e^{2 i \phi } \cosh ^2(r)+\sinh ^2(r)+\cosh (T)\right)}{T \left(\sinh ^2(r)+\left(\cosh (T)-e^{i \phi } \cosh (r)\right)^2\right)}\cr
&Q=\frac{i e^{-i (\phi -\theta )} \sinh (r) \sinh (T) (\cosh (T)-1)}{T \left(\sinh ^2(r)+\left(\cosh (T)-e^{i \phi } \cosh (r)\right)^2\right)}\cr}\;.
\Label{PQ}
 $$
 where
$$
T=\cosh ^{-1}(\cos (\phi ) \cosh (r))\;.\Label{PQ2}
 $$
Note that if $\cos(\phi)\cosh(r)<1$, $T$ turns out to be imaginary, namely
$T=i\,\cos^{-1}(\cos (\phi ) \cosh (r))$.

\bigskip 
\noindent \tit{Example 5.} We consider a two--mode Gaussian unitary given by a 
beam splitter followed by a Caves--Schumaker unitary, followed by a 
two--mode displacement.  The beam splitter is  
specified by the rotation matrix
\beq\nn 
\phi=\bm 0 & -i \beta\\
 i \beta & 0 \\\em
\quad,\quad \beta\in\M(R)
\eeq
and has the following Hamiltonian representation and symplectic matrix
\beq\nn 
\BH_{b}=\bm
 0 & i \beta & 0 & 0 \\
 -i \beta& 0 & 0 & 0 \\
 0 & 0 & 0 & -i \beta\\
 0 & 0 & i \beta & 0 \\
\em
\quad,\quad
\BS_b=\bm
 \cos\beta & \sin\beta & 0 & 0 \\
 -\sin\beta & \cos\beta & 0 & 0 \\
 0 & 0 & \cos\beta & \sin\beta \\
 0 & 0 & -\sin\beta & \cos\beta \\
\em
\eeq
The Caves--Schumacher unitary is  a squeezer specified by the squeeze matrix
\beq\nn 
z=\bm
 0 & e^{i \theta } r \\
 e^{i \theta } r & 0 \\
\em
\quad,\quad r,\theta\in\M(R)
\eeq
and has the following Hamiltonian representation and symplectic matrix
\beq\nn 
\BH_c=\bm
 0 & 0 & 0 & i e^{i \theta } r \\
 0 & 0 & i e^{i \theta } r & 0 \\
 0 & -i e^{-i \theta } r & 0 & 0 \\
 -i e^{-i \theta } r & 0 & 0 & 0 \\
\em
\quad,\quad
\BS_c=\bm
 \cosh r & 0 & 0 & e^{i \theta } \sinh r \\
 0 & \cosh r & e^{i \theta } \sinh r & 0 \\
 0 & e^{-i \theta } \sinh r & \cosh r & 0 \\
 e^{-i \theta } \sinh r & 0 & 0 & \cosh r \\
\em
\eeq
The global symplectic matrix is given by
\beq\nn 
\BS=\BS_c\BS_b=\bm
 \cos \beta  \cosh r & \cosh r \sin \beta  & e^{i \theta } \sin \beta  \sinh r & e^{i \theta } \cos \beta  \sinh r \\
 -\cosh r \sin \beta  & \cos \beta  \cosh r & e^{i \theta } \cos \beta  \sinh r & 
-e^{i \theta } \sin \beta  \sinh r \\
 e^{-i \theta } \sin \beta \sinh r & e^{-i \theta } \cos \beta  \sinh r & \cos \beta  \cosh r & \cosh r \sin \beta  \\
 e^{-i \theta } \cos \beta  \sinh r & -e^{-i \theta } \sin \beta  \sinh r & -\cosh r \sin \beta  & \cos \beta  \cosh r \\\em
\eeq

The problem is the evaluation of the matrix log given by (\ref{PQ12}). Since the
$4\times4$ matrix $\BS$ has not distinct eigenvalues we 
cannot use Sylvester's formula, but the matrix log can be evaluated via Jordan canonical form \cite{Higham}. Considering that $\BS$
has two distinct eigenvalues with multiplicity 2, specifically 
\beq\nn 
\mu_{1,2}=\cos (\beta ) \cosh (r)\mp \frac{1}{2} \sqrt{2 \cos ^2(\beta ) \cosh (2 r)+\cos (2 \beta )-3}
 \eeq
 the related Jordan form is
\beq\nn 
\BH=i\Omega\,\log\BS=i\Omega \sum_{m=0}^{3} d_m\,\BS^m
\eeq
where
\beq\nn 
d_{0}=\frac{-(\mu _1-\mu _2) \left[\mu _1^2+\mu _2^2\right]+(\mu _1-3 \mu _2) \mu _1^2 \log (\mu _2)+(3 \mu _1-\mu _2) \mu _2^2 \log (\mu _1)}{(\mu _1-\mu _2)^3}
 \eeq
\beq\nn 
d_{1}=\frac{(\mu _1-\mu _2) (\mu _1+\mu _2) \left[\mu _1^2+\mu _2 \mu _1+\mu _2^2\right]+6 \mu _1^2 \mu _2^2 (\log (\mu _2)-\log (\mu _1))}{\mu _1 (\mu _1-\mu _2)^3 \mu _2}
 \eeq
\beq\nn 
d_{2}=\frac{-2 \mu _1^3+2 \mu _2^3+3 \mu _2 (\mu _1+\mu _2) \mu _1 (\log (\mu _1)-\log (\mu _2))}{\mu _1 (\mu _1-\mu _2)^3 \mu _2}
 \eeq
\beq\nn 
d_{3}=\frac{\mu _1^2-\mu _2^2+2 \mu _2 \mu _1 (\log (\mu _2)-\log (\mu _1))}{\mu _1 (\mu _1-\mu _2)^3 \mu _2}
 \eeq
 Finally, we  evaluate the matrix $\Psi=(\BS-I_4)\BH^{-1}$, but
we find  a very complicated expression. However, we are interested in the evaluation of the coefficients $P$ and $Q$ for which we have obtained
a readable expression. We let
\beq
 M=
\sqrt{ \left[e^{2 r}+1\right]^2 \cos ^2(\beta )-4 e^{2 r}}=
=e^r\sqrt{2 \cos (2 \beta ) \cosh ^2(r)+\cosh (2 r)-3}
\nn\eeq
\beq
a=2 \log \left[\cos (\beta ) \cosh (r)-\frac{1}{2} M e^{-r}\right]
\nn\eeq
\beq
b=2 \log \left[\cos (\beta ) \cosh (r)+\frac{1}{2} M e^{-r }\right]
\nn\eeq
\beq
c=\log \left[2 \cos (\beta ) \cosh (r)-M e^{-r}\right]
\nn\eeq
\beq
d=\log \left[2 \cos (\beta ) \cosh (r)+M e^{-r}\right]
\nn\eeq
 \beq
h=e^{i \theta }\sqrt{ \left[e^{2 r}+1\right]^2 \cos (2 \beta )-6 e^{2 r}+e^{4 r}+1}
\nn\eeq
\beq\nn 
k=\log 4 +2r\log \left[ \left[e^{2 r}+1\right] \cos (\beta )+M  \right]
\nn\eeq
\beq
m=\log 4+2 r\log \left[ \left[e^{2 r}+1\right] \cos (\beta )-M  \right]
\nn\eeq
Then we get
\begin{eqnarray}
&P&=\frac{i \left[a+b\right] (\cos\beta -1)+2 i \sqrt{2}\left[e^{2 r}+1\right] 
e^{i \theta } \sin ^2\beta (c-d)}{h k m} \\
&Q&=\frac{i \left[a+b\right] \sin\beta -2 i \sqrt{2} \left[e^{2 r}+1\right] e^{i \theta } \sin\beta 
(\cos\beta -1) (c-d )}{h k  m}
\end{eqnarray}

\bigskip

\section{Amplification of the displacement}
\bigskip
Starting from a linear Hamiltonian  $H=H_\ell=\xi^*\Bh$
we can generate a displacement whose amount is just given by $\Bh$. But,
introducing a Gaussian unitary containing rotation and squeezing 
we can modify 
the displacement amount as 
\beq\nn 
   \Bs=\Psi\,\Bh\;.
\eeq
  
In particular, acting on the rotation and squeezing parameters we can obtain an 
amplification of the displacement, as seen for the single and the two--mode.
In the following consideration  we first consider  the case 
squeezing+displacement and then we consider the general case where
 also the rotation is present.

\subsection{The displacement amount with only squeezing}

We have seen in Fig.3 and in Fig.4 that in the presence of only squeezing we get both 
attenuation and amplification, where a fundamental role is played by the squeeze 
phase $\theta$.  In fact, the displacement amount is given by
(\ref{DD2FB}), that is, 
\beq
s=-i\frac{\sinh r}{r}\;h + 
	    i\frac{\cosh r-1}{r}e^{i\theta}\; \ov h\
	    \Label{PP2}
\eeq
For convenience we consider the case $h=1$,  so that the displacement amount 
is given by
\beq\nn
|\Bs(r,\theta)|= 
         \frac{\left|\sinh r+e^{i\theta }(\cosh r-1)\right|}{r}
=\frac{2 \sinh(r/2)\sqrt{\cosh r-\sinh r \cos\theta }}{r}\;.
\eeq

The function $|\Bs(r,\theta)|$, illustrated in \Figt(DE4013), has a maximum at 
$\theta=\pi$ \, given by $(e^r-1)/{r}$ and  a minimum at $\theta=0$, given
by $e^{-r}(e^r-1)/r$. The maximum and the minimum are as illustrated in 
\Figb(DD4025) (left) as a function of $r$.

In the $(r,\theta)$--plane the separation between  the amplification and
attenuation regions is determined by the condition
$$
|\Bs(r,\theta)|=\frac{2 \sinh(r/2) \sqrt{\cosh (r)-\sinh (r) \cos (\theta )}}{r}=1
$$
as illustrated in Fig. 11 (right).

\subsection{The displacement amount in the general case}

The presence of only rotation does not provide amplification 
but only attenuation. In  fact,
with $h=1$ we have the amount (see (\ref{DD9}))
\beq
 |s|=	\left|\frac{\sin(\phi/2)}{\phi/2}\right|\leq1
         \quad,\quad\phi\in[-\pi,+\pi)\;.     \Label{HH13a}
\eeq
 However, the rotation when combined with squeezing, provides
 huge amplifications, as already illustrated in Fig. 6, where the curves
 of $|s|$ versus $\phi$ exhibit divergences.
 We reconsider the plots  around the divergences
\beq\nn 
\phi_r=\arccos(-1/\cosh(r))\vq 2\pi-\phi_r\;.
\eeq
For instance, with $r=0.5$ we find $\phi_r=2.66121$.

 The conclusion seems that, combining squeezing and rotation,
 one can achieve arbitrarily huge displacement amounts! 

\section{Conclusions} 

\bigskip
Starting form the matrix representation $(\BH,\Bh)$ of the Hamiltonian 
 we have derived the
matrix representation $ (\BS,\Bs)$ of the complex symplectic transformation,
with the direct passage from the bosonic Hilbert space to the phase space.
This approach has 
 several advantages with respect to the traditional one, based
of Bogoliubov transformation and real symplectic transformations, as
illustrated in the introduction.  The derivation was carried out in the general
$n$--mode arriving in any case at explicit closed--form results.

 In particular, we have focused our attention
on the relation between the linear terms $\Bh$ and $\Bs$, not developed in the literature of Gaussian unitaries. This relation becomes $\Bs=\Bh$ 
in the presence of displacement only, that is, with $\BH=0$. In all the other cases
$\Bs$  turns out to be strongly dependent on the quadratic part $\BH$ of
the Hamiltonian.  As illustrated with the application to combination of  fundamental Gaussian unitaries,  from a given $\Bh\neq0$, it is possible to achieve an $\Bs$ arbitrarily large.  In the authors' opinion this topic
deserves a further development with the help of an experimental verification.

\bigskip

\ET {\Large \bf Acknowledgment}
\medskip
\ET The authors thank Gerardo Adesso and Antony  Lee
for having inspired the topic leading to Theorem 1 and for their useful cooperation.

\vspace{8mm}
{\Large\bf \quad APPENDIX}

\bigskip

{\large\bf Appendix A: Sum of the series $\Psi$ when $\BH$ is singular}
\bigskip

The series $\Psi$ defined by (\ref{Psi}) can be summed in a closed form
using the identity
\beq
\sum_{n=1}^\infty\frac1{n!}\,C^{\;n-1}=(e^{\,C}-I_p)C^{-1} \Label{B2K}
\eeq
where $C$ is a nonsingular $p\times p$ matrix.
If the matrix $\B(H)$ is not singular, the application of (\ref{B2K})
 to the series $\Psi$ gives the result written in (\ref{Psi}).

The series  can be summed in closed form also when $\B(H)$ is singular, using
the Jordan decomposition of the matrix $-i\Omega \B(H)$,  say
\beq\nn 
-i\Omega \B(H)=V \Lambda\,V^{-1}\,. \Label{CC1}
\eeq
If $r<2n$ is the rank of $\B(H)$ we decompose the diagonal matrix $\Lambda$ in
the form
\beq\nn 
  \Lambda=\bm\Lambda_r&0\\0&\Lambda_0\\\em \Label{CC2}
\eeq
where $\Lambda_r$ contains the Jordan blocks corresponding to the non vanishing
eigenvalues of $-i\Omega\B(H)$ e $\Lambda_0$ contains the Jordan blocks
corresponding to the zero eigenvalues. Then the series gives
\beq
\Psi=\sum_{n=1}^\infty\frac1{n!}(-i \Omega \B(H))^{n-1}(-i\Omega )
=V\sum_{n=1}^\infty \frac1{n!}
\bm\Lambda_r^{n-1}&0\\0&\Lambda_0^{n-1}\\\em V^{-1}(-i\Omega )
\label{A12L}
\eeq
Since $\Lambda_r$ is non singular (\ref{B2K}) gives
\beq
W_r=\sum_{n=1}^\infty \frac1{n!}\Lambda_r^{n-1}=(e^{\Lambda_r}-1)\Lambda_r^{-1}\;.\Label{CC4}
\eeq
Moreover, $\Lambda_0$ is nilpotent so that the series
\beq\nn 
W_0=\sum_{n=1}^\infty \frac1{n!}\Lambda_0^{n-1} \Label{CC6}
\eeq
reduces to a finite sum. In conclusion:

\medskip\noindent{\sc Proposition 3.} \tit{If $r<2n$ is the rank of $\B(H)$,
in the Jordan decomposition $-i\Omega\BH=V\Ld V^{-1}$, the matrix $\Ld$ is
 decomposed as
in (\ref{CC2}), where $\Ld_r$ is regular and $\Ld_0$ is nilpotent. Then
\beq 
\Psi=-i\,V\bm W_r&0\\0&W_0\\\em V^{-1}(-i\Omega) \Label{CC9}
\eeq
where $W_r$ is given by (\ref{CC4}) and $W_0$ by (\ref{CC6}).}

\bigskip

\mn\tit{Example 6 (Single mode)}. In the single mode  the matrix $\BH$ has the structure
\beq\nn 
\B(H)=\bm\alpha&\beta\\\beta^*&\alpha\\\em\;.
\eeq
Its singularity implies $|\beta|=|\alpha|$ and $\beta=\alpha e^{i\phi}$, so that
one gets
\beq\nn 
-i\Omega\B(H)=-i\alpha\bm 1&e^{i\phi}\\-e^{-i\phi}&-1\\\em\;.
\eeq
The matrix $-i\Omega\B(H)$ has a double eigenvalue $\lb=0$. On the other hand, it
cannot be diagonalizable because in this case it should vanish. On the contrary,
it is nilpotent since $(-i\Omega\B(H))^2=0$. As a consequence, it follows
\beq\nn 
\B(S)=e^{-i\Omega\B(H)}=I-i\Omega\B(H)=\bm 1-i\alpha&-i\alpha e^{i\phi}\\i\alpha
	e^{-i\phi}&1+i\alpha\\\em
\eeq
\beq\nn 
\Psi=\sum_{n=1}^\infty\frac1{n!}(-i\Omega \B(H))^{n-1}(-i\Omega )
=\left(I-\frac12i\Omega\B(H)\right)(-i\Omega)
%=\bm1-i\alpha/2&-i\alpha e^{i\phi}/2\\i\alpha
%	e^{-i\phi}/2&1+i\alpha/2\\\em\bm-i&0\\0&i\\\em
%\eeq
%\beq
=\bm-\frac\alpha2-i&\frac12\alpha e^{i\phi}\\\frac12\alpha e^{-i\phi}&-
\frac\alpha2+i\\\em\;.
\eeq

\medskip

\noindent\tit{Example 7 (A two mode).}  We consider a degenerate two--mode rotation 
specified by the Hermitian matrix
\beq\nn 
\phi=\bm\phi&0\\0&\phi^\TT\\\em\quad{\rm with}\quad \phi= 
\bm
 \phi _{11} & \phi _{12}  \\
 \phi _{12}^* & \phi _{22}  \\\em
\eeq
 The matrix representation of the Hamiltonian is
\beq\nn 
\BH=-\bm
 \phi _{11} & \phi _{12} & 0 & 0 \\
 \phi _{12}^* & \phi _{22} & 0 & 0 \\
 0 & 0 & \phi _{11} & \phi _{12}^* \\
 0 & 0 & \phi _{12} & \phi _{22} \\\em
\eeq
We have
$\det \BH=(|\phi _{12}|^2- \phi _{11}\phi _{22})^2
$, so that a degenerate case is obtained with
\beq\nn 
|\phi _{12}|=\sqrt{\phi _{11}\phi _{22}}\quad\to\quad {\rm rank}\,\BH=2\;.
\eeq
In this case we have
\beq
-i\Omega\BH=-i\bm\,
 \phi _{11} & \sqrt{\phi _{11} \phi _{22}} & 0 & 0 \\
 \sqrt{\phi _{11} \phi _{22}} & \phi _{22} & 0 & 0 \\
 0 & 0 & -\phi _{11} & -\sqrt{\phi _{11} \phi _{22}} \\
 0 & 0 & -\sqrt{\phi _{11} \phi _{22}} & -\phi _{22} \\\em
\Label{B17A}
\eeq
and  $r={\rm rank}(\BH)=2$. The matrices of the Jordan decomposition are 
\beq\nn 
V=\bm
 0 & -\frac{\phi _{22}}{\sqrt{\phi _{11} \phi _{22}}} & 0 & \frac{\phi _{11}}{\sqrt{\phi _{11} \phi _{22}}} \\
 0 & 1 & 0 & 1 \\
 -\frac{\phi _{22}}{\sqrt{\phi _{11} \phi _{22}}} & 0 & \frac{\phi _{11}}{\sqrt{\phi _{11} \phi _{22}}} & 0 \\
 1 & 0 & 1 & 0 \\\em
\quad,\quad
\Lambda=\bm
 0 & 0 & 0 & 0 \\
 0 & 0 & 0 & 0 \\
 0 & 0 & -i (\phi _{11}+\phi _{22}) & 0 \\
 0 & 0 & 0 & i (\phi _{11}+\phi _{22}) \\\em
\eeq
\beq\nn 
V^{-1}=\bm
 0 & 0 & -\frac{\sqrt{\phi _{11} \phi _{22}}}{\phi _{11}+\phi _{22}} & \frac{\phi _{11}}{\phi _{11}+\phi _{22}} \\
	-\frac{\sqrt{\phi _{11} \phi _{22}}}{\phi _{11}+\phi _{22}} & \frac{\phi _{11}}{\phi _{11}+\phi _{22}} & 0 & 0 \\
	0 & 0 & \frac{\sqrt{\phi _{11} \phi _{22}}}{\phi _{11}+\phi _{22}} & \frac{\phi _{22}}{\phi _{11}+\phi _{22}} \\
	\frac{\sqrt{\phi _{11} \phi _{22}}}{\phi _{11}+\phi _{22}} & \frac{\phi _{22}}{\phi _{11}+\phi _{22}} & 0 & 0 \\\em
\eeq
\beq\nn 
\Lambda_r=\bm-i(\phi_{11}+\phi_{22})&0\\
                 0&i(\phi_{11}+\phi_{22})\\\em\vq
\Lambda_0=\bm0&0\\0&0\\\em\;.	 
\eeq
Hence
\beq\nn 
\sum_{n=1}^\infty \frac1{n!}\Lambda_r^n=(e^{\Lambda_r}-I_r)\Lambda_r^{-1}
=\bm
 \frac{i \left[-1+e^{-i (\phi _{11}+\phi _{22})}\right]}{\phi _{11}+\phi _{22}} & 0 \\
 0 & -\frac{i \left[-1+e^{i (\phi _{11}+\phi _{22})}\right]}{\phi _{11}+\phi _{22}} \\\em\vq\q W_0=0\;.
\eeq

\bigskip

{\large\bf Appendix B: Hamiltonian representation of the fundamental unitaries}

\bigskip\noindent
In the literature the multimode fundamental Gaussian unitaries (displacement,
rotation, and squeezing) are usually expressed in terms of the bosonic vectors
$a$ and $a_*$ (see f.i. \cite{MaRh90}). Then, it may be useful to find the
relations between these expressions and the corresponding quantities $\B(H)$ and
$\B(h)$.

The usual representation of a $n$--modal {\it displacement} operator is given by
\beq
D(\alpha)=e^{\alpha^\T a_*-\alpha^*a}\quad,\quad
\alpha=[\alpha_1,\ldots,\alpha_n]^\T\in\M(C)^n\;,\Label{dalpha}
\eeq
while its representation in terms of the Hamiltonian is (see (\ref{U22)}))
\beq\nn 
e^{-iH_\ell}=e^{-i\B(h)^*\xi}=e^{-ih^* a-ih^\T a_*}
\eeq
coinciding with (\ref{dalpha}), provided that $\B(H)=0$ and
$\B(h)=\bm i\alpha\\-i\ov\alpha\\\em$.

The {rotation} operator is usually given by (see (\ref{U24}))
\beq\nn 
R(\phi)=e^{ia^*\phi a}\;,\label{rphi}
\eeq
where $\phi$ is a Hermitian $n\times n$ matrix. The present representation
of the rotation is given by the Hamiltonian
\beq\nn 
H_r={1\over2}\sum_{r,s=1}^n[\,A_{rs}a_r^*a_s+\overline A_{rs}a_ra_s^*\,]
=\frac12\sum_{r,s=1}^n[\,A_{rs}a_r^*a_s+A_{sr}(a_s^*a_r-\delta_{rs})\,]=
a^*Aa-\frac12{\rm Tr}(A)
\eeq
so that, apart an irrelevant phasor,
\beq\nn 
e^{-iH_r}=e^{-ia^*Aa}
\eeq
coinciding with (\ref{rphi}) provided that $\B(h)=0$ and $A=-\phi$, i.e.,
$\B(H)=\bm-\phi&0\\0&-\overline\phi\\\em$.

Finally, the {\it squeezing} operator is given by (see (\ref{U26}))
\beq\nn 
Z(z)=e^{{1\over2}[a^*za_*-a\overline za]}\;,\label{zz}
\eeq
with $z$ a symmetric $n\times n$ matrix. The Hamiltonian is given by
\beq\nn 
H_s={1\over2}\sum_{r,s=1}^n[\,B_{rs}a_r^*a_s^*+\overline B_{rs}a_ra_s\,]
={1\over 2}(a^*Ba_*+a^T\overline Ba)
\eeq
so that
\beq\nn 
e^{-iH_s}=e^{-{\over 2}}(a^*Ba_*+a^T\overline Ba)
\eeq
coincides with (\ref{zz}), provided that $\B(h)=0$ and $B=iz$, i.e,
$\B(H)=\bm 0&iz\\-i\overline z&0\\\em$.

The expressions of the FUs in terms of the vectors $a,a_\star$ (traditional
  form)  and in terms of the single vector $\xi$ are summarized 
  in the following table:
\bigskip
 
 \begin{tabular}{ll ll}
 \hline
  \multicolumn{4}{l}{\bf Displacement}\\
  $\q D(\alpha):=e^{\alpha^T a_*-\alpha^* a}$&\q $\alpha= \in\M(C)^n$
  &\qq    $\C(D)_Y(\xi)= e^{\xi^*\;\Omega\,Y}$&\qq$Y=\bm\alpha\\\alpha_* \\\em$\\
  \hline
  \multicolumn{4}{l}{\bf Rotation}\\
  $\q R(\phi):=e^{i\,a^*\phi \, a}$&\quad $\phi
 \quad n\times n$  Hermitian
 &\qq$\C(R)_\Phi(\xi)=e^{i\,\xi^*\,\Phi\,\xi}$&\quad \quad$\Phi=\bm\phi&0\\0&0\\\em$\\
 \hline
  \multicolumn{4}{l}{\bf Squeezing}\\
  $\q Z(z):=e^{\frac12\left[\,( a^* \,z\,a_*-a^\TT\,z^*\,a)\right]}$ 
&\q $z$\q $n\times n$ symmetric
&\qq$ \C(Z)_Z(\xi)=e^{i\frac12\xi^*\;Z\;\xi}$ &\quad\q
               $ Z=\bm0& z\\
	                       z^*&0\\\em$\\
\hline			       
 \end{tabular}
\bigskip

\bigskip
{\large\bf Appendix C: Proof of Propositions 1 and 2}
\bigskip

For Proposition 1, starting from
\beq\nn 
{\bf-i\Omega\B(H)}=\bm i\phi&0\\0&-i\phi^\TT\\\em
\eeq
one finds immediately
\beq\nn 
\B(S)=\bm e^{i\phi}&0\\0&e^{-i\phi^\T}\\\em\;.
\eeq Then, from (\ref{Psi}) it follows
\beq\nn 
\Psi=(\B(S)-I_{2n})\B(H)^{-1}=-\bm(e^{i\phi}-I_n)\phi^{-1}&0\\0&(e^{-i\phi^\T}-I_n)(\phi^{-1})^\TT\em\;.
\eeq
For Proposition 2, starting from (\ref{DD2A}) and using the polar decomposition
$z=re^{i\theta}$ gives
\beq\nn 
-i\Omega\B(H)=\bm0&re^{i\theta}\\re^{-i\theta^\T}&0\\\em
\eeq
As shown in \cite{MaRh90}, the corresponding Bogoliubov transformation gives
\beq\nn 
e^{i\Omega\B(H)}\bm a\\ a_*\\\em e^{-i\Omega\B(H)}=\bm \cosh(r)a+\sinh(r)e^{i\theta}a_*\\
\cosh(r^\T)a_*+\sinh(r^\T)e^{-i\theta^\T}\em\;,
\eeq
so that the complex symplectic matrix is
\beq\nn 
\B(S)=\bm\cosh(r)&\sinh(r)e^{i\theta}\\\sinh(r^\T)e^{-i\theta^\T}&\cosh(r^\T)\em\;.
\eeq
In conclusion, from (\ref{Psi}) one gets
\beq\nn 
\Psi=\bm\cosh(r)-I_n&\sinh(r)e^{i\theta}\\\sinh(r^\T)e^{-i\theta^\T}&\cosh(r^\T)-I_n\em
\bm0&i\ov z^{-1}\\-iz^{\,-1}&0\\\em\;,
\eeq
from which (\ref{HH4}) follows.

\bigskip
\newpage
{\large\bf Appendix D: Alternative evaluation of $\Psi$ in the single mode} 

\bigskip\noindent

We consider the matrix $\BH$ in the single mode
\beq
\BH=\left[
\begin{array}{cc}
 a & b \\
 b^* & a \\
\end{array}\right]\vq\q a>0\vl b\in\M(C)
 \eeq
and we evaluate the corresponding  symplectic matrix $\BS$. We find
\beq
\BS=e^{-i\Omega \BH}=\left[
\begin{array}{cc}
 \cosh (T)-\frac{i a \sinh (T)}{T} & -\frac{i b \sinh (T)}{T} \\
 \frac{i b^* \sinh (T)}{T} & \cosh (T)+\frac{i a \sinh (T)}{T} \\
\end{array}\right]  \Label{ZA4}
 \eeq
 where
 \beq\nn
T=\sqrt{|b|^2-a^2}\vq T^2=|b|^2-a^2\;.
 \eeq
On the other hand we know that the   symplectic matrix matrix has the form
\beq
\BS=\bm\cosh(r)\E^{\I\phi}&\sinh(r)\E^{\I(\theta-\phi)}\\
             \sinh(r)\E^{-\I(\theta-\phi)}&\cosh(r)\E^{-\I\phi}\\\em
	    \Label{ZA6}
	     \eeq
Now {\bf we assume to know the parameters $r,\theta,\phi$ and 
we want to evaluate
the parameters $a,b$.}  To this end  we
equate the first rows  of (\ref{ZA4}) and (\ref{ZA6})
\beq
\cosh (T)-\frac{i a \sinh (T)}{T}=\cosh(r)\E^{\I\phi}\vq
 -\frac{i b \sinh (T)}{T}=
      \sinh(r)\E^{\I(\theta-\phi)} \Label{ZA8}
 \eeq 
  Assuming as known $T$, the solution is
 \beq\FB{
 a=-i T \frac{\cosh (T)-e^{i \phi } \cosh (r)}{ \sinh(T)}
 \vq b= i T\frac{\sinh (r) e^{i \theta -i \phi }}  {\sinh(T)}
 } 
\eeq
 To calculate $T$ we take the real part of the first of (\ref{ZA8})
 \beq\nn
 \cosh(T)=\cosh(r)\cos(\phi)\q\to\q T= \cosh^{-1}[\cosh(r)\cos(\phi)]
 \eeq
 Hence 
 \beq\nn
 \FB{ T= \cosh^{-1}[\cosh(r)\cos(\phi)]}
 \eeq
and
\beq\nn \sinh(T)=\sqrt{\cosh(T)^2-1}
=\sqrt{\cosh^2(r)\cos^2(\phi)-1}
\eeq
This complete the evaluation of $\BH$ from $\BS$.

Once evaluated the Hamiltonian matrix in terms of the parameters
$r,\theta,\phi$, we can calculated the matrix $\Psi= (\BS-I_2)\BH^{-1}$
 as a function of the same parameters. We find
\beq\nn
\Psi=\left[
\begin{array}{cc}
 -\frac{i \tanh (T) \left(-e^{i \phi } \cosh (r) (\cosh (T)+1)+e^{2 i \phi } \cosh ^2(r)+\sinh ^2(r)+\cosh (T)\right)}{T \left(\sinh ^2(r)+\left(\cosh (T)-e^{i \phi } \cosh (r)\right)^2\right)} & \frac{i e^{-i (\phi -\theta )} \sinh (r) (\cosh (T)-1) \tanh (T)}{T \left(\sinh ^2(r)+\left(\cosh (T)-e^{i \phi } \cosh (r)\right)^2\right)} \\
 \frac{e^{-i \theta } \sinh (r) \tanh (T) (\sin (\phi )-i \cos (\phi )) (2 \cos (\phi ) \cosh (r)-\cosh (T)-1)}{T \left(\sinh ^2(r)+\left(\cosh (T)-e^{i \phi } \cosh (r)\right)^2\right)} & \frac{i e^{-i \phi } \tanh (T) \left(\cosh (r) \left(\cosh (T)+e^{2 i \phi }\right)-e^{i \phi } (\cosh (T)+1)\right)}{T \left(\sinh ^2(r)+\left(\cosh (T)-e^{i \phi } \cosh (r)\right)^2\right)} \\
\end{array}\right]
 \eeq
 In particular
\beq\nn
P=-\frac{i \sinh (T) \left(-e^{i \phi } \cosh (r) (\cosh (T)+1)+e^{2 i \phi } \cosh ^2(r)+\sinh ^2(r)+\cosh (T)\right)}{T \left(\sinh ^2(r)+\left(\cosh (T)-e^{i \phi } \cosh (r)\right)^2\right)}
 \eeq
\beq\nn
Q=\frac{i e^{-i (\phi -\theta )} \sinh (r) \sinh (T) (\cosh (T)-1)}{T \left(\sinh ^2(r)+\left(\cosh (T)-e^{i \phi } \cosh (r)\right)^2\right)}
 \eeq
 where
\beq\nn
\cosh(T)=\cos (\phi ) \cosh (r)
 \eeq
\beq\nn
\sinh(T)=\sqrt{\cos ^2(\phi ) \cosh ^2(r)-1}\;.
 \eeq


\begin{thebibliography}{longlonglong}
\bibitem{spec} The equivalence can be proved by showing 
that it is possible to pass from  one 
specification to the other \cite{CarFU}.
\bibitem{squeeze} The squeeze operator is
             usually denoted by the letter $S$, but this is in conflict
	      with the notation used for the symplectic matrix 
	      encountered in symplectic transformations.
\end{thebibliography}
\end{document}